\documentclass[12pt]{article}
\usepackage{epsfig}
\begin{document}

\begin {center}
{\Large \bf Light $2^{++}$ and $0^{++}$ mesons}
\vskip 5mm
{$^a$A.V. Anisovich , $^b$D V Bugg, $^a$V.A. Nikonov,
$^a$A.V. Sarantsev and  $^a$V.V. Sarantsev \\
{\normalsize $^a$ \it PNPI, Gatchina, St. Petersburg district, 188350,
Russia}\\
{\normalsize $^b$ \it Queen Mary, University of London, London
E1\,4NS, UK}
\\ [3mm]}
\end {center}
\date{\today}

\begin{abstract}
The status of light $I=0$, $J^{PC}=2^{++}$ and $0^{++}$ mesons is
discussed, particularly the separation of $n\bar n$ and $s\bar s$
states.
They fall into a simple scheme except for $f_2(1810)$.
A case is made that this has been confused with the $f_0(1790)$.
It should be possible to check this with existing or forthcoming data.  

\vskip 2mm
{\small PACS numbers: 14.40.Be, 14.40.Df, 13.75.-n}
\end {abstract}

\section {Introduction}
The objective of this paper is a critical review of
the identification of light mesons with $J^{PC}$ = $2^{++}$ and $0^{++}$,
particularly the separation of $2^{++}$ states into
$\bar nn$ and $s\bar s$.
That detail is important in its own right. 
It will also be vital input information into the search for the
$2^{++}$ glueball in radiative $J/\psi$ decays.
It is unrealistic that those data will be
extensive enough to sort out the complex spectroscopy of
$q\bar q$ states as well as glueballs.

Section 2 reviews $2^{++}$ states. They fall into a regular
scheme except for $f_2(1810)$, whose identification is
presently weak.
Section 3 reviews $0^{++}$ states and makes the case that
$f_2(1810)$ may have been confused with $f_0(1790)$.
Without an $f_0$ in that mass range, the $0^{++}$ spectrum
is obviously incomplete, but $f_0(1790)$ fits in naturally.
Section 4 draws conclusions and comments on improvements needed in 
present partial wave analyses to resolve this question; these are
straightforward and can be done with existing data and/or
forthcoming BES III data.

\section {$J^P = 2^{++}$ states}
There are extensive data from Crystal Barrel on $\bar pp$
scattering in flight to 17 final states.
These concern purely neutral final states which can be classified
into four families with isospin $I=0$ and 1, and charge 
conjugation $C = +1$ and $-1$.
We shall concentrate here on $I=0$, $C = +1$ where the data
are most complete.
This is a formation experiment of the type $\bar pp \to resonance \to A + B$.
These data are listed by the PDG under `Further States',
requiring confirmation \cite {PDG}.
That is not possible for most states because other data come from
production reactions of the form  
$\pi p \to X + p$ (or n), where the exchanged meson is uncertain, 
leading to ambiguities in partial wave analysis; also they have
no polarisation data.

A combined analysis has been published \cite {combined} of 8 sets of 
data with $I=0$, $C= +1$ on 
$\bar pp \to \pi^0\pi ^0$, $\eta \eta$, $\eta \eta '$, and $\eta \pi^0
\pi ^0$ from Crystal Barrel at 9 beam momenta, plus measurements of 
differential cross sections and polarisation from the PS172 experiment 
\cite {PS172} and an earlier experiment of Eisenhandler et al. at the 
CERN PS \cite {Eisenhandler}.
The first set of $\pi^+\pi ^-$ data covers the momentum range 0.36--1.55 
GeV/c and the second covers 1--2.5 GeV/c.
They agree accurately where they overlap.
The data of PS172 are particularly valuable because they extend down
to a beam momentum of 360 MeV/c (a mass of 1910 MeV) in quite small
steps of momentum (30--100 MeV/c) and therefore cover
in detail the lower side of the cluster of resonances around 2000 MeV; 
they also used a beam going through the detector and therefore cover 
centre of mass scattering angles to $\cos \theta > 0.999$.
Two further analyses were reported of Crystal Barrel data for 
$\bar pp \to \eta ' \pi ^0\pi ^0$ \cite {eprpp} and $3\eta$
\cite {3eta}.
They find masses and widths consistent within errors with the
combined analysis. 

The polarisation data are very important.
Because $^3P_2$ and $^3F_2$ have orthogonal Clebsch-Gordan coefficients,
they separate those partial waves accurately.
Differential cross sections contain real parts of interferences
within singlet and triplet sets of amplitudes; polarisation data
measure the imaginary part of interferences within the
triplet set. 
Formulae are given in Ref. \cite {formulae}.
Column 5 of Table 1 gives the ratio $r_J = g_{L=J+1}/g_{L=J-1}$ of
amplitudes, where $g_L$ are coupling constants for
$\bar pp$ orbital angular momentum $L$; intensites depend on
$r_J^2$.
The phase sensitivity of the polarisation data improves greatly
the accuracy of masses and widths. 
The only singlet states with $I=0$, $C=+1$ are $^1S_0$, $^1D_2$ and $^1G_4$ 
and these states are separated by their angular dependence.

The analysis relies on fitting with analytic functions of $s$ and assumes
two towers of states in mass ranges 1910-2100  and 2200-2370 MeV; it includes
Blatt-Weisskopf centrifugal barrier factors with a radius which optimises at
$0.83 \pm 0.021$ fm. 
The tails of $\eta _2(1870)$ and $f_6(2465)$ are included using masses and
widths determined elsewhere. 
It is fortunate that $^3F_4$ states near 2050 and 2300 MeV are strong and
accurately determined by their rapid angular dependence, 
and act as powerful interferometers to determine lower partial
waves. 
Starting from these partial waves and adding lower $J^P$, the analysis
finds a unique set of amplitudes; only in two low partial waves with little 
or no angular dependence are there sizable errors in fitted masses and
widths.
Recently, as a convenience, we have installed the relevant publications
on the arXiv system, and give references in the bibliography. 
Further details and figures of data are given in a full length review 
\cite {review}.
There is a total of $>10$ million fully reconstructed events.
The data and Monte Carlo sets are publicly available from the authors
of this paper, subject to a joint publication of results. 
A complete set of data is also available on $\bar pp$ annihilation at 
rest in liquid hydrogen and deuterium and in gas for both.
This makes one of the largest data sets available in meson spectroscopy.

\begin{table}[htp]
\begin{center}
\begin{tabular}{ccccccc}
\hline
State & Mass & Width & $s\bar s$ Mixing & $r_J =$ & Observed \\
      &$(MeV/c^2)$ & $(MeV/c^2)$ &Angle (deg) &
$\frac {g(L=J+1)}{g(L=J-1)}$ & Channels \\
\\\hline
$f_2(1270)$ [9] & $1270 \pm 8$ & $194 \pm 36$ & 0 & 0 & $\pi
\pi$,$4\pi$ \\
$f_2(1565)$ [9] & $1560 \pm 15$ & $280 \pm 40$ & 0 & 0 & $\pi
\pi$,$\eta \eta$,$\omega\omega$ \\
$f_2(1910)$ [2,7] & $1934 \pm 20$ & $271 \pm 25$ & 1.1 & $0.0 \pm 0.08$
& $\pi \pi$,$\eta \eta$,$f_2\eta$,$a_2\pi$ \\
$f_2(2000)$ [2,7] & $2001 \pm 10$ & $312 \pm 32$ & 7.9 & $5.0 \pm 0.5$
& $\pi \pi$,$\eta \eta$,$\eta \eta '$, $f_2\eta $\\
$f_2(2240)$ [2,7] & $2240 \pm 15$ & $241 \pm 30$ & 7.5 & $0.46 \pm 0.09$
& $\pi \pi$,$\eta \eta$,$\eta \eta '$, $f_2\eta $\\
$f_2(2295)$ [2,7] & $2293 \pm 13$ & $216 \pm 37$ &-14.8 & $-2.2 \pm 0.6$
& $\pi \pi$,$\eta \eta$,$\eta \eta '$, $f_2\eta $,$a_2\pi$\\\hline
$f_2(1525)$ [10] & $1513 \pm 4$ & $76 \pm 6$ & & &$K^+K^-$ \\
$f_2(1525)$ [11] & $1508 \pm 9$ & $79 \pm 8$ & & &$\eta \eta$ \\
$f_2(1755)$ [12] & $1755 \pm 10$ &$67 \pm 12$ & & &$K^+K^-$ \\
$f_2(2150)$ [1]  & $2157 \pm 12$ & $152 \pm 30$ & & &$\eta \eta$,
$K_S^0\bar K_S^0$ \\
$f_2(2300)$ [13] & $2297 \pm 28$ & $149 \pm 41$ & & &
$K\bar K$, $\phi \phi$ \\
$f_2(2340)$ [13] & $2339 \pm 55$ & $319 ^{+81}_{-69}$ & & &
$\eta \eta $, $\phi \phi$ \\\hline
\end {tabular}
\caption {$I=0$ $J^{PC} = 2^{++}$ Resonances primarily discussed here,
$n\bar n$ in the top half of the table, $s\bar s$ in the bottom half.}
\end{center}
\end{table}
Table 1 lists $I=0$, $J^{PC}=2^{++}$ states. 
The top half of the table lists states which are dominantly $n\bar n$
and the lower half $s\bar s$ states.
Masses and widths are from Crystal Barrel where available.

There is also evidence for a broad $f_2$ listed by the PDG as
$f_2(1950)$; in Crystal Barrel data it appears in the $\eta \eta$
channel with a mass of $2010 \pm 25$ MeV and a width of $495 \pm 35$
MeV.
It is observed by other groups in $\pi \pi$, $\eta \eta$, $4\pi$, $KK$ and 
$KK\pi \pi$. 
It is a candidate for the $2^+$ glueball.
Another possibility is that it is a dynamically generated state related
to the opening of the strong $4\pi$ and $KK\pi \pi$ thresholds. 

\subsection {Separation of $n\bar n$ and $s\bar s$ states}
There is a publication concerning Crystal Barrel data in
flight which determines the mixing angle between $n\bar n$ and $s\bar
s$ using data on $\bar pp \to \pi ^0 \pi ^0$, $\eta \eta$ and $\eta
\eta '$ \cite {formulae}.
To our knowledge, these are the only data making a clean identification
of $n\bar n$ and $s\bar s$ and their mixing.
Resonances $R$ are linear combinations:
\newpage
\begin {equation}
R = \cos \Phi (|u\bar u> + d\bar d>)/\sqrt {2} + \sin \Phi |s\bar s>.
\end {equation}
The quark content of $\eta $ and $\eta '$ may be written in terms of
the pseudoscalar mixing angle $\Theta$ as
\begin {eqnarray}
|n\bar n> &=& \cos \Theta |\eta> + \sin \Theta |\eta '>, \\
|s\bar s> &=& -\sin \Theta |\eta> + \cos \Theta |\eta '>.
\end {eqnarray}
Amplitudes for decays of $R$ are given by
\begin {eqnarray}
f(\pi ^0 \pi ^0) &=& \cos \Phi /\sqrt {2}, \\
f(\eta \eta) &=& \cos \Phi (\cos ^2 \Theta + \sqrt {2\lambda}
\sin ^2\Theta \tan  \Phi)/\sqrt {2}, \\
f(\eta \eta ') &=& \cos \Phi \cos \Theta \sin \Theta
(1 - \sqrt {2\lambda} \tan \Phi )\sqrt {2}.
\end {eqnarray}
where $\sin \Theta$ was taken as 0.6 and $\lambda $ as 0.85.
Values of the $\bar ss$ mixing angle $\Phi$  are listed in column 4 of
Table 1.
Errors are not listed in the original publication, but are quoted as
typically $\pm 5^\circ$.
Each mixing angle of Table 1 is consistent with zero within
three standard devations, though from the overall $\chi^2$ there is a
a definite indication that some small mixing with $s\bar s$ occurs;
that is to be expected from $\bar ss$ states across the mass range. 
The conclusion is that none of the states listed in the
first half of the table is dominantly $s\bar s$.
Amongst recognised or possible $s\bar s$ states, a small signal from
$f_2(1525)$ can be detected in Crystal Barrel data for $K\bar K$ and
$\eta \eta$ via its interference with $f_0(1500)$ \cite {Amsler02},
\cite {Anisee}.
The $f_0(980)$ appears as a dip in the Dalitz plot for
$\bar pp \to 3\pi^0$, but with a much smaller branching ratio compared
with the $\sigma$ amplitude than in $\pi \pi$ elastic scattering,
see Figs 2 and 4(a) of Ref. \cite {3pizero}.

Otherwise, recognised $s\bar s$ states are conspicuous by
their absence in $\bar pp$ annihilation.
The $f_2(2300)$ and $f_2(2340)$ are observed by Etkin et al. in 
$K\bar K$ and $\phi \phi$  \cite {etkin}. 
The initial state for those data is $\pi \pi$, but no $\pi \pi$ decays 
are observed, showing that any $\pi\pi$ coupling in these states, 
hence $n\bar n$ component, must be  small. 
Presently, the PDG lists $f_2(2240)$ under $f_2(2300)$.
That is clearly inconsistent with the mixing angles of Table 1
and needs to be corrected in PDG tables. 
The $f_2(2295)$ is missing from the tables and needs to be
included, since it is observed in five channels of data: 
$\pi \pi$, $\eta \eta$, $\eta \eta '$, $f_2(1270)\eta$ and
$a_2(1320)\eta$; dropping it from the $\pi \pi$ channel 
alone increases $\chi^2$ by 2879, which is highly significant.
Figures 9(n) and 9(o) of that paper illustrate the 
effects of dropping them from the analysis and re-optimising
all other components; the changes near $\cos \theta = 1$ are
very large, leaving no doubt of their significance.

Visual evidence for $f_2(2240)$ and $f_2(2295)$ are also displayed
in the first analysis of $\bar pp \to \eta \pi ^0\pi ^0$ data in Ref. \cite {epp}.
Fig. 16 of that paper displays the requirement for 
$f_2(2240) \to [f_2(1270)\eta]_{L=1}$, where $L$ is the angular momentum
in the decay.
In the final combined analysis of Ref. \cite {combined}, log likelihood
is worse by 468 if $f_2(2240)$ is omitted in the $\eta \pi ^0 \pi ^0$ channel
and by 1557 if $f_2(2295)$ is omitted; here log likelihood is defined so
that it increases by 0.5 for a one standard deviation change in each
coupling constant.

There is no convincing evidence in Crystal Barrel data for $f_2(2150)$,
which is observed by other groups only in $\eta \eta$ and $K\bar K$.
The $f_2(2150)$ may be interpreted as the $s\bar s$
$^3P_2$ partner of $f_2(1910)$.
Their mass difference is similar to that between $f_2(1270)$ and
$f_2(1525)$.
The $f_2(2300)$ and $f_2(2340)$ are observed by Etkin et al. in
$K\bar K$ and $\phi \phi $ S and D-waves.
Both $f_2(2300)$ and $f_2(2340)$ may be interpreted as the
partner of the $^3F_2$ $n\bar n$ state at 2001 MeV, with decays to
$\phi \phi$ S and D-waves; the L=2 dependence makes the
D-wave peak higher.
However, the 150 MeV mass gap between $f_2(2150)$ and $f_2(2300)$
is a little surprising compared with the 90 MeV gap between $f_2(1910)$ and
$f_2(2000)$.

Under $f_2(2150)$, the PDG  lists Anisovich 99K
data on $\bar pp \to \eta \eta \pi ^0$ \cite {Anisee}
as presenting evidence that there is a state at $2105 \pm 10$ MeV
consistent with the $f_2(2150)$.
In fact, the paper presented a careful study of both the angular
distribution, which is flat, and the energy dependence of
production (which is different for production of $J=2$ and 0); the  
conclusion from both sources is that the signal is due to $f_0(2105)$, 
which is observed in many sets of data and unambiguously has $J^P=0^+$. 
It is the only state observed in Crystal Barrel data with a large 
mixing angle $(68-71.6)^\circ$ to $s\bar s$.
It makes up $(4.6 \pm 1.5)\%$ of the $\pi ^0\pi^0$ intensity and
$(38 \pm 5)\%$ of $\eta \eta$.
The branching ratio to $\eta \eta'$ is not well determined because of
low statistics in this channel.
The best estimate of amplitude ratios is
\begin {equation}
\pi ^0 \pi ^0:\eta \eta:\eta \eta ' =
0.71 \pm 0.17:1:-0.85\pm 0.45.
\end {equation}
For an unmixed $q\bar q$ state, the ratio expected between $\pi ^0\pi
^0$ and $\eta \eta$ is $0.8^{-4} = 2.44$.
A possible interpretation is that it is an $s\bar s$ state mixed with
$f_0(2020)$.
An alternative is that it is the second $0^+$ glueball predicted by
Morningstar and Peardon in this general mass range \cite {glueball}.
A pointer in this direction is that it was first identified in
Mark III data for $J/\psi \to \gamma (4\pi)$ \cite {MarkIII}.
A pure glueball would have a mixing angle of $+37^\circ$.
Its strong coupling to $\bar pp$ is clearly anomalous.

We are able to check the partial wave analyses of
$\bar pp \to \pi ^+\pi ^-$ done by Hasan et al. \cite {Hasan} and Oakden and
Pennington \cite {Oakden}.
Because their analyses were limited to this channel alone, errors on
mass and width are larger than from the full analyis by a factor
$\ge 5$.
This explains why their results have larger errors and fluctuations.
Our experience is that, as a rule of thumb, each data set
with large statistics reduces errors by a factor 2 because different
sets of data have different sensitivity to details.
The PDG attributes the $f_2(2226)$ reported by Hasan  \cite
{Hasan} as $f_2(2150)$.
That is a mistake.
It should be attributed to $f_2(2240)$.

 \begin{figure}[htb]
 \begin{center}
 \vskip -11.5mm
 \epsfig{file=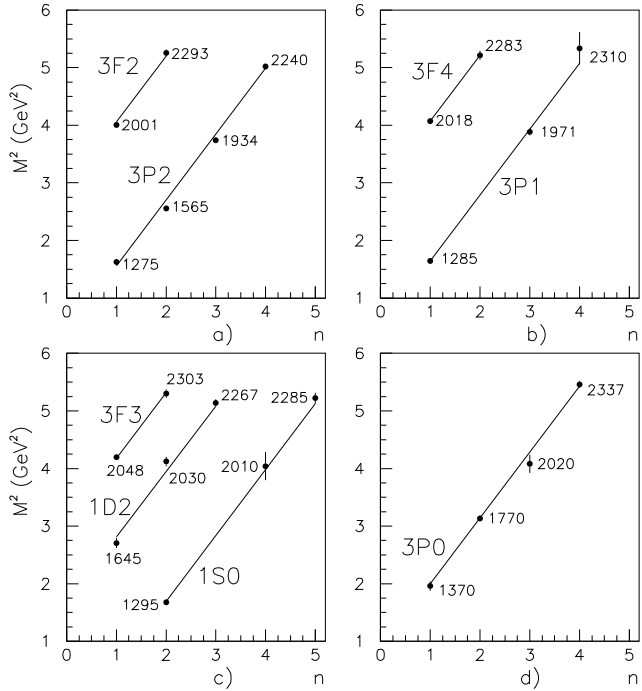,width=10.5cm}
 \vskip -5.5mm
 \caption {Trajectories of light mesons with $I=0$, $C=+1$ observed in
 Crystal Barrel data in flight, plotted against radial excitation number
$n$; masses are marked in MeV. }
 \end{center}
 \end{figure}

The outcome of this analysis is that states observed in the Crystal Barrel 
analysis for the four sets of quantum numbers $I=0$, $C=\pm 1$ and 
$I=1$, $C = -1$ fall on to parallel trajectories displayed in Figs. 1 
and 3 of Ref. \cite {review}.
Those with $I=0$, $C = +1$ are shown here in Fig. 1.
They are particularly well identified because of the availability of
the polarisation data.
Some states are significantly displaced from straight line
trajectories by thresholds.
A striking example is $f_2(1565)$ which coincides with the $\omega
\omega $ threshold and is displaced downwards from its isospin partner
$a_2(1700)$ by $\sim 135$ MeV.
The origin of this shift is a narrow cusp in the real part of
the amplitude at the opening of any sharp threshold, as explained in
Ref. \cite {sync}.

The experimental data separate $\bar pp$ $^3F_2$ and $^3P_2$ states.
These can be interpreted as $^3F_2$ and $^3P_2$ $q\bar q$ configurations 
for the following reasons.
A feature of the data is that $F$-states decay strongly to
channels with high orbital angular momentum.
The origin of this effect is clearly a good overlap between initial
and final state wave functions.
Llanes-Estrada et al. point out a formal analogy with the Frank-Condon
principle of molecular physics consistent with this observation
\cite{Estrada}.
In essence, this analogy provides a mechanism via which
$n\bar n$ $^3F_2$ states couple preferentially to $\bar pp$ $^3F_2$
states and likewise $^3P_2$ $n\bar n$ states couple preferentially
to $\bar pp$ $^3P_2$ states.

It is observed that $^3F_2$ states lie systematically higher in mass
than $^3P_2$ by $\sim 60$ MeV, see Fig. 1.
For $I=1$, $C = -1$, D-states lie roughly midway in mass.
These observations are consistent with stronger centrifugal
barriers in $F$-states delaying the appearance of $F$-wave
resonances to higher masses.
The spin-splitting between $F$-states is consistent within errors with
tensor splitting, which is predicted to be dominantly from one-gluon
exchange between $q\bar q$ \cite {Godfrey}.
The $^3F_4$ states have some admixtures of $^3H_4$; the $^3H_6$
state does not appear until $2465 \pm 60$ MeV \cite {PDG}.
The overall picture is the first appearance of the lowest-lying
$F$-states at $\sim 2030$ MeV. 
This is what is to be expected from Regge trajectories for $n\bar n$ 
states.

\subsection {Discussion of other $J^{PC} = 2^{++}$ listings of the PDG.}
There are other candidates for $2^{++}$ states listed by the PDG.
Firstly, the $f_2(1640)$ has a simple explanation.
It is the $\omega \omega $ decay mode of $f_2(1565)$ \cite {Baker}.
The latter state sits precisely at the $\omega \omega$ threshold.
The $f_2(1640)$ has a line-shape which is well fitted by folding
$\omega \omega$ S-wave phase space with the line-shape of $f_2(1565)$.
It is fitted like that by Baker et al., including the dispersive term
which originates from the opening of the $\omega \omega$ threshold.
The square of the coupling constant to $\rho \rho$ is 3 times that
for $\omega \omega$ by SU(2) symmetry. 
The result is that $\rho \rho$ decays of $f_2(1565)$ also peak
at $\sim 1640$ MeV, but the peak is broader than for $\omega \omega$
when the widths of the two $\rho$ are folded in, see Fig. 5(b) of
Ref. \cite {Baker}. 

The $f_2(1430)$ listed by the PDG has an explanation, illustrated in 
Fig. 10 of Ref. \cite {1370}.
Below the $\omega \omega$ threshold, the Breit-Wigner denominator of
$f_2(1565)$ needs to include an analytic continuation of the phase
space factors below both $\rho \rho$ and $\omega \omega$ thresholds.
The analytic continuation causes a phase variation in $\pi \pi$ and leads to
an interference between the $f_2(1565)$ and $f_2(1270)$.
This  interference is very clear for the $\pi \pi $ D-wave in Crystal
Barrel data for $\bar pp$ at rest $\to 3\pi ^0$.
The effect is maximal at a mass of 1420 MeV.
Similar interferences in other channels listed under $f_2(1430)$ may
be explained this way.
It was a good observation by the experimental groups before the
existence of the $f_2(1565)$ was well known.

The $f_2(2010)$ listed by the PDG has a simple explanation.
The peak observed in the data of Etkin et al. \cite {etkin} is
at 2150 MeV and agrees with the $f_2(2150)$.
It is not necessary to have two separate $f_2$ states at 2010 and 2150
MeV.
The partial wave analysis of Etkin et al. used the K-matrix
approach.
It is possible to have a K-matrix pole at the $\phi \phi$ threshold,
significantly displaced from the T-matrix pole at 2150 MeV.

\section {Discussion of $f_2(1810)$ and $J^P = 0^+$ states}
This leaves the $f_2(1810)$.
It does not fit well on to the $2^+$ trajectories shown in Fig. 1;
it appears to be an `extra' state.
Dudek \cite {dudek} has recently presented a lattice QCD calculation
of hybrid masses and light mesons.
He predicts a lowest group of hybrids with $J^{PC}= 1^{-+}, 0^{-+}$,
$1^{--}$ and $2^{-+}$.
His mass scale needs to be normalised against the well known $f_4(2050)$
and the $\rho (1700)$ $^3D_1$ states.
It then agrees quite well with the exotic $\pi_1(1600)$ (actually at
1660 MeV), the $\pi (1800)$ and the two `extra' $2^{-+}$ states $\eta_2(1870)$ 
and $\pi_2(1880)$, which do not fit into the $^1D_2$ trajectory of Fig. 1(c) 
above.  
No $2^{++}$ hybrid is predicted in this mass range. 

This prompts a careful re-examination of the data on which $f_2(1810)$ is based.
The $f_2(1810)$ is not well established.
Also there is clear evidence for an $\bar nn$ $J^{PC}=0^{++}$
state $f_0(1790)$ very close to this mass, distinct from $f_0(1710)$.
We shall consider possible confusion between $J=0$ and 2 for this
state.

The primary data for $f_2(1810)$ come from the GAMS collaboration
\cite {GAMS1}.
These are mostly on $\pi ^-p \to 4\pi ^0 n$.
In these data, the separation between $J^P = 2^+$ and $0^+$
rests on the number of events observed below and above
$\cos \theta = 0.4$, where $\theta$ is defined in their paper.
In practice this is a rather fine distinction.
The argument is that $f_2$ events are enhanced at small $\cos
\theta$ and $f_0$ events are enhanced at large $\cos \theta$.
That is counter-intuitive and must depend strongly on the
Monte Carlo of acceptance which is quoted but not shown.
It would be valuable if the Compass collaboration could check
these results in $\pi ^+\pi ^-\pi ^+\pi ^0$  and in
$4\pi ^0$ if that is possible.

Data of Costa et al. \cite {Costa} are also quoted, but these
refer to $\pi ^-p \to K^+K^-n$.
There are eight alternative solutions, which mostly contain some mild
peaking near 1800 MeV.
These could be explained by the $f_2(1755)$ of the L3 collaboration.
The PDG quotes also an $f_2(1857^{+18}_{-71})$ fitted by Longacre
to data available in 1986 \cite {Longacre}.
At that time, $f_2(1910)$ had not been discovered.
If it had been known at the time, there would undoubtedly have been
some perturbation to Longacre's analysis.
Finally, the observation of a structure at 1799 MeV in $\pi ^+p \to
\Delta ^{++} \pi ^0\pi ^0$ by Cason et al. was not confirmed by
Prokoshkin et al \cite {Prokoshkin}.
Cason et al. argue that their charge exchange data choose a unique
solution from four ambiguous solutions to $\pi ^+\pi ^-$ elastic
scattering.
It is however somewhat puzzling that their solution does not
contain any significant signal for $f_0(1500)$, which ought to be
conspicuous.
Neither do they observe the $f_2(1565)$ which is conspicuous in the
$\pi \pi$ channel; indeed, that is where it was first observed by
the Asterix collaboration \cite {May}.

The alternative assignment for the 1810 MeV state is $J^P=0^{++}$.
Here, important data come from BES II for $J/\psi$ decays.
There are six relevant sets of data.
The first two are for $J/\psi \to \omega K^+K^-$ \cite {wKK} and
$\omega \pi ^+\pi ^-$ \cite {wpp}.
A large $f_0(1710)$ signal is observed in $\omega K^+K^-$.
In contrast, high statistics data on $\omega \pi^+\pi ^-$ show no
structure in $\pi^+\pi ^-$ at this mass.
Those data set an upper limit of $11 \%$ on the branching ratio
$\pi \pi / K\bar K$ of $f_0(1710)$ with $95\%$ confidence.

The third and fourth sets of data are for
$J/\psi \to \phi \pi ^+ \pi ^-$ and $\phi K^+K^-$ \cite {phipp}.
In the $\phi \pi ^+\pi ^-$ data there is a definite $J^{PC}=0^{++}$
$\pi \pi$ peak, but at $1790 ^{+40}_{-30}$ MeV, visibly distinct from
the mass of $f_0(1710)$.
There is no definite evidence for a decay to $K\bar K$, though a small
amount can be fitted.
There is a factor 22--25 discrepancy between the branching ratio
$\pi \pi /K\bar K$ for these data and the data for $J/\psi \to \omega
\pi \pi$ and $\omega KK$, so this peak cannot be due to $f_0(1710)$.
The discrepancy points strongly to the existence of a second
$\bar nn$ state distinct from $f_0(1790)$.
 \begin{figure}[htb]
 \begin{center}
 \vskip -10mm
 \epsfig{file=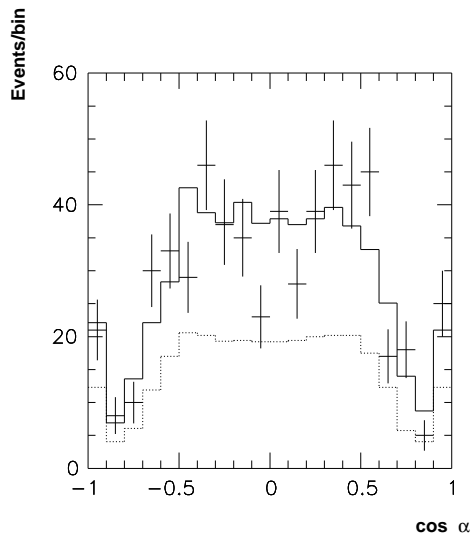,width=8cm}
\vskip -3mm
 \caption {Angular distribution for BES II data on $J/\psi \to \phi \pi ^+\pi ^-$
(points with errors); the full histogram shows the fit with $f_0(1790)$
and the dotted line the acceptance. }
 \end{center}
 \end{figure}

A question is whether these data can alternatively be fitted by
$f_2(1810)$.
The observed decay angular distribution, shown here in Fig. 2, is flat 
within experimental errors in the region where the acceptance of the
BES II detector (dotted histogram) is uniform. 
However, the acceptance falls rapidly at $|\cos \theta| \simeq 0.6$.
The BES publication comments that the $f_2(1810)$ can be produced with 
orbital angular momentum $\ell = 0$, 2 or 4 in the production step.
The $\ell = 0$ component is likely to be dominant and gives a
decay angular distribution proportional to the Legendre polynomial
$P_2(\alpha) = 3\cos ^2 \alpha - 1$, where $\alpha$ is the decay angle 
of the $\pi ^+$ from $f_J(1790)$ in the resonance rest frame.
On resonance, the $\phi$ and $f_J(1790)$ are produced with momenta
of 630 MeV/c in the lab frame, and the centrifugal barrier reduces
$\ell = 2$ amplitudes by a factor $\sim 2$ and the $\ell = 4$
amplitude by a factor $\sim 30$, so the $\ell = 4$ amplitude may
safely be neglected.
For $\ell = 2$, there are three combinations of $\ell = 2$ with 
spin $j=2$ of the $f_2(1790)$, making total spins $S=0$, 1 and 2.
Of these, spin 1 makes two amplitudes  proportional to $\sin^2 \alpha$
and $\sin \alpha \cos \alpha$, but no amplitude proportional to $P_2$.
The $S=0$ amplitude is proportional to $P_2(\alpha)$.
The $S=2$ amplitude is formed from $S=2$ and the spin 2 of $f_2(1790)$
and contains a $P_2$ term.
The publication says: `Our experience elsewhere is that using four
helicity amplitudes instead of two adds considerable flexibility to the fit.
We conclude that the state is most likely spin zero.'

\begin{table}[htp]
\begin{center}
\begin{tabular}{ccc}
\hline
Entry & Amplitudes & Change in log likelihood \\\hline
A &  $\ell =0$ & -216 \\
  & $\ell = 2$, $S=0$         & -467 \\
  & $\ell = 2$, $S=1$         & -475 \\
  & $\ell = 2$, $S=2$         & -254 \\
B & $\ell=0+\ell=2,S=1$       & -35 \\
  & $\ell=0+\ell=2,S=2$       & -135 \\
  & $\ell=0+\ell=2,S=0$       & -184 \\
C & $\ell=0+\ell=2,S=1$ and 2 & -4 \\\hline
\end {tabular}
\caption {Changes in log likelihood with a variety of $f_2(1790)$ amplitudes
fitted to BES II data on $J/\psi \to \phi \pi \pi$ and $\phi KK$.}
\end{center}
\end{table}
It is now worth amplifying this comment with numbers in Table 2 from  
the analysis; these may be understood in terms of the acceptance. 
Each amplitude is fitted freely in magnitude and phase.
Angular correlations with the decay of the $\phi$ are included.
The fit to $f_0(1790)$ produced with both $\ell = 0$ and 2 amplitudes
is taken as a benchmark. 
Further entries in the table show changes in log likelihood (defined
so that a change of +1 is better by one standard deviation for
two degrees of freedom).
The $\ell = 2$ amplitude changes sign at $|\cos \alpha | \simeq 0.577$.
In entry A, changes in log likelihood are shown for four single
$f_2$ amplitudes.
Entry B shows the best three pairs of $f_2$ amplitudes and Entry C the
best combination of 3.

In A, the best fit with $\ell = 0$ is considerably worse than the
benchmark, but uses the $P_2(\alpha)$ dependence to produce a
fit peaking at $\cos \alpha = 0$ and dropping sharply at $|\cos \alpha | = 0.6$, 
though it gives a false peak near $|\cos \alpha | = 1$.
The fit with $S=0$ requires strong correlations between the
production angle and decay angle and is considerably worse.
The fit with $S=1$ is bad because two of the contributions go
to zero in the middle of the angular distribution.

In B, the best fit is a combination of $L=0$ and $L=2$ with $S=1$.
The second of these helps produce the box-shaped distribution of 
data on Fig. 2 and is able to compensate $P_2$ to some extent near
$|\cos \alpha | = 1$.
In C, three amplitudes can produce nearly as good a fit as $f_0(1790)$.
There is considerably flexibility using three fitted phase angles;
random phases give much worse fits.

It is of course possible that $f_2(1810)$ is produced with a set of
amplitudes which happen to agree with $f_0(1790)$. 
Further BES III data with considerably improved acceptance and higher
statistics have a high chance of resolving the situation.

\subsection {X(1812)}
The BES II collaboration also present data on
$J/\psi \to \gamma (\omega \phi)$ \cite {BESwphi}.
There is a clear peak in $\omega \phi$ at $1812 ^{+19}_{-26} \pm 18$
MeV.
Quantum numbers $J^{PC} = 0^{++}$ are favoured quite
significantly over $2^{++}$ and $0^{-+}$.
The $\omega \phi$ channel opens at 1802 MeV.
The present data may be fitted within the sizable errors by folding 
the line shape of $f_0(1790)$ with $\omega \phi$ phase space, using a 
reasonable form factor $\exp (-2k^2)$, where $k$ is the momentum in the 
$\omega \phi$ channel in GeV/c.
The PDG lists this state under $X(1835)$,  observed in
$J/\psi \to \gamma (\eta '\pi ^+\pi ^-)$ \cite {X1835A}
\cite {confirmed}.
However, as the PDG remarks, $J^P=0^+$ is not allowed for this
final state.
The angular distribution of the photon for those data is consistent
with $J^P = 0^-$, but might be accomodated with $1^{++}$ if the
helicity ratio of the two possible $1^{++}$ amplitudes is just right.
But neither of these possibilities is consistent with the observed
peak presently attributed to $f_0(1790)$.

The $X(1812)$ decays to $\phi \omega$.
There is an important simplification that in radiative production of
$X(1812)$ there are only three $J^P=2^+$ helicity amplitudes instead of
five because helicity 0 is forbidden for the photon.
An analysis of the spin correlation between these two
would identify $J^P$ of $X(1812)$.
The spin of the $\phi$ is measured by $(p_1 - p_2)$,
where $p_1$ and $p_2$ are momenta of the kaons from its decay in
its rest frame; the spin of the $\omega$ is normal to the decay
plane of the $\omega$ in its rest frame.
For spin 0, the angular distribution of $\phi$ and $\omega$ decays
is given by the dot product of these two vectors.
This is a delicate test of the spin of the $X(1812)$.
Formulae for other $J^P$ are given by Zou and Bugg \cite{ZouBugg}.

The decay to $\phi \omega$ is surprising  (Okubo-Zweig-Iizuka violating).
It could arise from a glueball component mixed into $X(1812)$.
Two gluons couple to 
$(u\bar u + d\bar d + s\bar s)(u\bar u + d\bar d + s\bar s)$.
The cross-terms between $u\bar u + d\bar d$ and $s\bar s$ can generate
$\omega \phi$.

The $f_0(1790/1812)$ would fit naturally on to the $0^{++}$ trajectory
of Fig. 1.
It would not be surprising that there are two $0^+$ states close
in mass; a similar pair is $f_2(1525)$ and $f_2(1560)$.
There is earlier independent evidence for an $f_0(1750)$ in Mark III data for
$J/\psi \to \gamma 4\pi$ \cite {MarkIII}.
There is a further observation of a well defined $0^{++}$
signal in $\eta \eta$ in Crystal Barrel data in flight at 
$1770 \pm 12$ MeV with width $220 \pm 40$ MeV \cite {CBARee}.
The mass is 4 standard deviations above $f_0(1710)$ and
the width is 2 standard deviations higher.
This signal could come from a superposition of $f_0(1790)$ with 
$f_0(1710)$, which clearly has a large $s\bar s$ component (and/or glueball).

\section {Conclusions}
We have made a case that existing light mesons with $J^P=2^+$
fall into a regular pattern of $n\bar n$ $^3P_2$ and $^3F_2$
states except for $f_2(1810)$.
There is the possibility that it has been confused with
$f_0(1790)$; if not, there is a missing $0^+$ state on the
trajectory of Fig. 1 at a similar mass.
The pattern of $f_0(1710)$ and $f_0(1790)$ is like that of
$f_2(1525)$ and $f_2(1565)$.

We remark that the search for the $2^{++}$ glueball will
require full use of existing identifications of $q\bar q$
and $s\bar s$ components with these quantum numbers.
We have pointed out some corrections to PDG Tables.

We also remark that, in the long term, it would be possible to do 
further polarisation measurements in $\bar pp$ scattering in the 
beam momentum range from $\sim 360$ to 1940 MeV/c at the 
forthcoming FAIR facility \cite {further}. 
This was part of the proposed program at LEAR, but was cut
short by the closure of that machine. 
Measurements of polarisation in $\bar pp \to \eta \pi ^0\pi ^0$
and $\eta '\pi ^0\pi ^0$ are realistic and would give
information on interferences between singlet and triplet
partial waves.
For $I=1$, $C=+1$ and $I=0$, $C=-1$, there are presently no
polarisation data. 
Such data would improve vastly the identification of states
with these quantum numbers.

\end {document}